\documentclass[aps,pra,twocolumn,amsfonts,amssymb,amsmath,showpacs,
floatfix,nofootinbib,citesort]{revtex4-2}
\usepackage{mathrsfs}
\usepackage{amsfonts}
\usepackage{amstext}
\usepackage{amsmath}
\usepackage{amssymb}
\usepackage{bm}
\usepackage{CJK}
\usepackage{bbm}
\usepackage[dvips]{graphicx}
\def\qed{\leavevmode\unskip\penalty9999 \hbox{}\nobreak\hfill
     \quad\hbox{\leavevmode  \hbox to.77778em{%
              \hfil\vrule   \vbox to.675em%
               {\hrule width.6em\vfil\hrule}\vrule\hfil}}
     \par\vskip3pt}

\usepackage{amssymb}
\usepackage{graphicx}
\usepackage{graphics}
\usepackage{amsmath}
\usepackage{amsthm}
\usepackage{color}
\usepackage{dsfont}
\usepackage{textcomp}
\definecolor{darkred}  {rgb}{0.5,0,0}
\definecolor{darkblue} {rgb}{0,0,0.5}
\definecolor{darkgreen}{rgb}{0,0.5,0}
\usepackage{hyperref}
\hypersetup{
	pdftitle = {QRT Proposal},
	pdfauthor = {},
	colorlinks = true,
	urlcolor  = blue,         
	linkcolor = red,     
	citecolor = blue,    
	filecolor = darkred       
}
\usepackage{mathtools}
\def\ra{\rangle}
\def\la{\langle}

\def\ot{\otimes}
\newtheorem{theorem}{Theorem}

\newtheorem{pro}{Proposition}

\newcommand{\bea}{\begin{eqnarray}}
\newcommand{\eea}{\end{eqnarray}}
\newcommand{\be}{\begin{equation}}
\newcommand{\ee}{\end{equation}}
\newcommand{\ba}{\begin{equation}\begin{aligned}}
\newcommand{\ea}{\end{aligned}\end{equation}}
\newcommand{\rank}{\text{Rank}}

\newcommand{\beax}{\begin{eqnarray*}}
\newcommand{\eeax}{\end{eqnarray*}}
\newcommand{\bex}{\begin{equation*}}
\newcommand{\eex}{\end{equation*}}

\theoremstyle{remark}

\def\be{\begin{equation}}
\def\ee{\end{equation}}

\newcommand{\mH}{\mathcal{H}}

\newcommand{\mS}{\mathcal{S}}

\newcommand{\tr}{{\rm Tr}}

\def\>{\rangle}
\def\<{\langle}

\begin{document}


\preprint{APS/123-QED}
\begin{CJK*}{GB}{gbsn}
\title{Complete monogamy of the multipartite quantum mutual information\\}


\author{Yu Guo}
\email{guoyu3@aliyun.com}
\author{Lizhong Huang}
\affiliation{Institute of Quantum Information Science, School of Mathematics and Statistics, Shanxi Datong University, Datong, Shanxi 037009, China}




\begin{abstract}
	Quantum mutual information (QMI) not only displays the mutual information in the system but also demonstrates some quantum correlation beyond entanglement. 
	We explore here the two alternatives of multipartite quantum mutual information (MQMI) based on the von Neumann entropy according to the framework of the \textit{complete} measure of multi-particle quantum system.
	We show that these two MQMI are complete, monogamous on pure states, 
	and one of them is not only \textit{completely monogamous} but also \textit{tightly complete monogamous} while another one is not.
	Moreover, we present another two MQMI by replacing the   
	von Neumann entropy with the Tasllis $q$-entropy from the former two ones.
	It is proved that one of them displays some degree of ``completeness'' as a measure of multi-particle quantum system, but the other one is not even non-negative and thus it can not be a alternative of MQMI.
	We also discuss the triangle relation for these three alternatives of MQMI.
	It is shown that the triangle inequalities hold for the former two MQMI as that of entanglement measure but the later one fails.
	By comparison, we found that the von Neumann entropy is better than other versions of entropy as desired when we characterize
	the quantum correlation in multi-particle system. 	
\end{abstract}


\maketitle
\end{CJK*}


\section{Introduction}

Quantum correlation, such as entanglement~\cite{Nielsen}, EPR (Einstein-Podolsky-Rosen) steering~\cite{Wiseman2007prl,Wiseman2007pra}, quantum discord~\cite{Ollivier2001quantum,Henderson2001classical}, etc., has been shown to be an essential resource to
achieve quantum advantages in various non-classical information processing tasks~\cite{Nielsen,Bennett1993teleporting,Zhang2006experimental,Bennett1992communication,Horodecki2009,Guhne2009,Bennett1996prl,Datta2008prl}.
One of the foremost 
issues in this area is to understand and quantify the
various forms of quantum correlations, especially for the multi-particle quantum system.
Consequently,
a series of multipartite entanglement measure~\cite{Verstraete2003pra,Luque2003pra,Osterloh2009jmp,Gour2010prl,Viehmann2011pra,Szalay}, multipartite quantum discord~\cite{Rulli2011pra,Giorgi2011prl,Radhakrishman2020prl}, multipartite quantum mutual information (MQMI)~\cite{Watanabe,Kumar2017pra}
have been proposed.

From the information-theoretical point of view, another crucial issue for multiparticle quantum system
is the distribution of the correlation up to the given measure.
The first contribution in this connection is the
monogamy relation of entanglement~\cite{Terhal2004}, which states
that, unlike classical correlations, if two parties $A$ and $B$ are
maximally entangled, then neither of them can share entanglement
with a third party $C$. Entanglement monogamy has many applications not only in quantum physics~\cite{Bennett2014,Toner,Seevinck} but also
in other area of physics, such as no-signaling theories~\cite{streltsov2012are,Augusiak2014pra}, 
condensed matter physics~\cite{Ma2011,Brandao2013,Garcia}, 
statistical mechanics~\cite{Bennett2014}, and even black-hole physics~\cite{Lloyd}.
Particularly, it is the key feature 
that guarantees quantum key distribution secure~\cite{Terhal2004,Pawlowski}.
The fundamental matter in this context is to
determine whether a given measure of quantum correlation is monogamous.
Indeed, intense research has been undertaken in this
direction. It has been proved that almost all the bipartite entanglement measures so far
are monogamous~\cite{Coffman,Osborne,streltsov2012are,
	Bai,Koashi,
	Luo2016pra,Dhar,Hehuan,GG2019,GG}.
However, these monogamy relations discussed via the bipartite measures (e.g., the entanglement measures) 
display certain drawback: only the
relation between $A|BC$, $AB$ and $AC$ are revealed, the global correlation
in $ABC$ and the correlation contained in part $BC$ is missed, where the vertical bar indicates 
the bipartite split across which we will measure the (bipartite) 
correlation. 
To address such a subject, the so-called \textit{complete monogamy} relation has been explored for multipartite measures~\cite{Guo2020pra,G2021qst,Guo2022entropy}.
In such a context, the multipartite measure should be \textit{complete}
in the sense that the correlation between any subsystem(s) with the coarsening relation could be compared with each other~\cite{Guo2020pra,G2021qst,Guo2022entropy}.

It has been showed that many complete multipartite entanglement measures are completely monogamous, i.e., any tripartite state (we take the tripartite case here) that satisfies~\cite{Guo2020pra}
\bea
E(ABC)=E(AB)
\eea 
implies $E(AC)=E(BC)=0$,
which is equivalent to
\bea
E^{\alpha}(ABC)\geqslant E^{\alpha}(AB)+E^{\alpha}(AC)+E^{\alpha}(BC)
\eea
holds for any state for some $\alpha>0$ whenever $E$ is continuous, where $E$ is a tripartite entanglement measure.
In Ref.~\cite{G2021qst}, with the same strategy as the complete multipartite entanglement measure established in Ref.~\cite{Guo2020pra}, the concept of complete multipartite quantum discord is investigated and is proved that
the mulitipartite quantum discord is completely monogamous if it is complete.

The mutual information, originally defined for classical system, the reciprocal information that is common to or shared by two or
more parties, has an authoritative stand in the arena of information theory. Quantum
mutual information is well defined for
bipartite quantum systems, i.e.,
\bea
I(A:B)&=&S(A)+S(B)-S(AB)\nonumber\\
&=&S(AB\|A\ot B)\geqslant0,
\eea
where $S(\rho)=-\tr(\rho\log_2\rho)$ is the von Neumann entropy
and $S(\rho\|\sigma)=\tr(\rho\log_2\rho-\rho\log_2\sigma)$ is the quantum relative
entropy, $S(X):=S(\rho^X)$, $\rho^{A,B}=\tr_{B,A}\rho^{AB}$ is the reduced state of $\rho^{AB}$. It reflects the total
correlation between the two subsystems~\cite{Groisman2005pra}. 
It can be generalized into multipartite case by different ways~\cite{Watanabe,Kumar2017pra}.
A natural 
$n$-party QMI defined in literature is
\bea\label{def-of-mui}
&&I(A_1:A_2:\cdots :A_n)\nonumber\\
&:=&\sum_{k=1}^nS({A_k})
-S(A_1A_2\cdots A_n)\nonumber\\
&=&S(A_1A_2\cdots A_n\|A_1\ot A_2\ot\cdots\ot A_n).\quad
\eea
Another alternative is~\cite{Kumar2017pra}
\bea\label{def-of-mui'}
I'(A_1:A_2:\cdots :A_n)\quad\quad\quad\quad\quad\quad\quad\quad\quad\quad\quad\quad\nonumber\\
:=\sum_{k=1}^nS(\overline{A_k})-(n-1)S(A_1A_2\cdots A_n)~\quad\quad\quad~~\nonumber\\
=S\left[ (A_1A_2\cdots A_n)^{\ot{n-1}}\|\overline{A_1}\ot \overline{A_2}\ot\cdots\ot \overline{A_n}\right],
\eea
where $\overline{X}$ denotes the complementary subsystems to that of $X$.
It is clear that $I\geqslant0$, $I'\geqslant0$, and that $I(A_1:A_2:\cdots :A_n)=0$ (or $I'(A_1:A_2:\cdots :A_n)=0$) if and only if 
$\rho^{A_1A_2\cdots A_n}$ is a product state since the relative entropy is non-negative and $S(\rho\|\sigma)=0$ iff $\rho=\sigma$. $I$ coincides with $I'$ for $n=2$ is the trivial case.
Any non-product state contains some quantum correlation~\cite{Guo2014srep,Guo2015ijtp}, and further more it is showed by the special three-qubit state that
the mutual information increases when the entanglement increases~\cite{Kumar2017pra}.
So the QMI displays some quantum correlation beyond entanglement in the system, and thus
it can be also regarded as a measure of some kind of quantum correlation.

The main purpose of this paper is to investigate whether the MQMI 
is a well-defined multipartite measure from the strategy in Ref.~\cite{Guo2020pra,Guo2022entropy,G2021qst}. Namely, whether the MQMI is complete, monogamous, and completely monogamous.
Throughout this paper, we let $\mH^{A_1A_2\cdots A_n}$
be an $n$-partite Hilbert space with finite dimension and let $\mS^{X}$ be the set of density
operators acting on $\mH^{X}$. $\rho^X$ (sometimes $\rho_X$) denotes the state in $\mS^{X}$.

The rest of this paper is arranged as follows.
In Sec. II, we review the notion of the coarser relation for multipartite partition
of multipartical state, which is convenient for discussing the complete measure of multipartite quantum correlation.
Sec. III discusses whether the MQMI $I$ and $I'$ are complete with the same spirit as the complete multipartite entanglement measure and the complete multipartite quantum discord put forward in literature.
In Sec. IV, we show that $I$ is completely monogamous but $I'$ is not, and that they are monogamous only one pure states. 
In Sec. V we establish the mutual information in terms of the Tsallis $q$-entropy and explore the complete monogamy  
accordingly. Furthermore, we explore the triangle inequality for these different MQMI and the relation with entanglement in Sec. VI.
Finally, we conclude with some discussions in Sec. VII.

\section{Coarser relation of multipartite partition}

We recall the coarser relation of multipartite partition proposed in Ref.~\cite{Guo2022entropy}.
Let $X_1|X_2| \cdots |X_{k}$ be a $k$-partition of $A_1A_2\cdots A_m$, i.e., $X_s=A_{s(1)}A_{s(2)}\cdots A_{s(f(s))}$, $s(i)<s(j)$ whenever $i<j$, and $s(p)<t(q)$ whenever $s<t$ for any possible $p$ and $q$, $1\leqslant s,t\leqslant k$.
For instance, partition $AB|C|DE$ is a $3$-partition of $ABCDE$.
Let $X_1|X_2| \cdots |X_{k}$ and $Y_1|Y_2| \cdots |Y_{l}$ be two partitions of $A_1A_2\cdots A_n$ or subsystem of $A_1A_2\cdots A_n$.
$Y_1|Y_2| \cdots |Y_{l}$ is called \textit{coarser} than $X_1|X_2| \cdots |X_{k}$~\cite{Guo2022entropy}, denoted by 
\bea
X_1|X_2| \cdots| X_{k}\succ Y_1|Y_2| \cdots |Y_{l}, 
\eea
if $Y_1|Y_2| \cdots |Y_{l}$ can be obtained from $X_1|X_2| \cdots| X_{k}$
by one or some of the following ways: 
\begin{itemize}
	\item[(a)] Discarding some subsystem(s) of $X_1|X_2| \cdots| X_{k}$;
	\item[(b)] Combining some subsystems of $X_1|X_2| \cdots| X_{k}$;
	\item[(c)] Discarding some subsystem(s) of some subsystem(s) $X_k$ provided that $X_{k}=A_{k(1)}A_{k(2)}\cdots A_{k(f(k))}$ with $f(k)\geqslant2$.
\end{itemize}
For example, $A|B|C|D|E \succ A|B|C|DE\succ A|B|C|D\succ AB|C|D\succ AB|CD$, $A|B|C|DE\succ A|B|DE$.

Furthermore, if $X_1|X_2| \cdots| X_{k}\succ Y_1|Y_2| \cdots |Y_{l}$,
we denote by $\Xi(X_1|X_2| \cdots| X_{k}- Y_1|Y_2| \cdots |Y_{l})$~\cite{Guo2022entropy} the set of
all the partitions that are coarser than $X_1|X_2| \cdots| X_{k}$ and either exclude any subsystem of $Y_1|Y_2| \cdots |Y_{l}$ or include some but not all subsystems of $Y_1|Y_2| \cdots |Y_{l}$.
We take the five-partite system $ABCDE$ for example,
$\Xi(A|B|CD|E-A|B)=\{CD|E$, $A|CD|E$, $B|CD|E$, $A|CD$, $B|CD$, $B|C|E$, $B|D|E$, $A|D|E$, $A|C|E$, $A|E$,
$B|E$, $A|C$, $A|D$, $B|C$, $B|D$, $C|E$, $D|E\}$.

For more clarity, we denote by~\cite{Guo2022entropy} 
\bea
X_1|X_2| \cdots| X_{k}\succ^a Y_1|Y_2| \cdots |Y_{l},\\
X_1|X_2| \cdots| X_{k}\succ^b Y_1|Y_2| \cdots |Y_{l},\\
X_1|X_2| \cdots| X_{k}\succ^c Y_1|Y_2| \cdots |Y_{l}~
\eea 
for the case of of (a), (b), and (c), respectively. 
For example, $A|B|C|D\succ^a A|B|D\succ^a B|D$,
$A|B|C|D\succ^b AC|B|D\succ^b AC|BD$, $A|BC\succ^c A|B$, $A|BC\succ^c A|C$.

\section{Completeness of mutual information}

Let $Q^{(n)}$ be a multipartite measure (for entanglement or quantum discord, ect). If $Q^{(k)}$ is uniquely determined by $Q^{(n)}$ for any $2\leqslant k<n$, then we call $Q^{(n)}$ a \textit{uniform} measure~\cite{G2021qst}. 
For example, the $n$-partite entanglement of formation~\cite{Guo2020pra},
is a uniform multipartite entanglement measure, which is defined as
$E_f^{(n)}( |\psi\rangle)=\frac12\sum_{i=1}^mS( \rho^{A_i})$
for pure state, where $\rho^{A_i}=\tr_{\overline{A_i}}|\psi\ra\la\psi|$, and via the convex-roof extension for mixed states (i.e.,
$E_{f}^{(n)}(\rho):=\min\sum_{i} p_iE_f^{(n)}(|\psi_i\ra)$
for any mixed state $\rho$, where the minimum is taken over all pure-state
decomposition $\{p_i,|\psi_i\ra\}$ of $\rho$).
It is clear that both $I$ and $I'$ are uniform measures.

In Ref.~\cite{Guo2020pra}, the complete multipartite entanglement measure is defined.
With the same spirit in mind, we discuss the completeness of the MQMI as a measure of
multipartite quantum system.
For more clarity, we recall the definition of the complete multipartite entanglement measure at first.
A uniform multipartite entanglement measure $E^{(n)}$ is called a \textit{unified}
multipartite entanglement measure if it also satisfies the~\textit{unification condition}~\cite{Guo2020pra}: 
i.e., 
	$E^{(n)}$ is consistent with $E^{(k)}$ for any $2\leqslant k<n$.
The unification condition should be comprehended in the following sense~\cite{Guo2020pra}:
\bea\label{additivity}
E^{(n)}(|\psi\ra^{A_1A_2\cdots A_k}|\psi\ra^{A_{k+1}\cdots A_n})\quad\quad\quad\quad\quad\quad\quad\quad\quad\nonumber\\
=E^{(k)}(|\psi\ra^{A_1A_2\cdots A_k})+E^{(n-k)}|\psi\ra^{A_{k+1}\cdots A_n},\quad
\eea
\bea
E^{(n)}(\rho^{A_1A_2\cdots A_n})=E^{(n)}(\rho^{A_{\pi(1)}A_{\pi(2)}\cdots A_{\pi(n)}})
\eea 
for any $\rho^{A_1A_2\cdots A_n}\in\mS^{A_1A_2\cdots A_n}$ and any permutation $\pi$, and
\bea\label{c1}
E^{(k)}(X_1|X_2| \cdots| X_{k})\geqslant E^{(l)}(Y_1|Y_2| \cdots |Y_{l})
\eea
for any $\rho^{A_1A_2\cdots A_n}\in\mS^{A_1A_2\cdots A_n}$ whenever $X_1|X_2| \cdots| X_{k}\succ^a Y_1|Y_2| \cdots |Y_{l}$,
where $X_1|X_2| \cdots |X_{k}$ and $Y_1|Y_2| \cdots |Y_{l}$ are two partitions of $A_1A_2\cdots A_n$ or subsystem of $A_1A_2\cdots A_n$.
A uniform $E^{(n)}$ is called a \textit{complete}
multipartite entanglement measure~\cite{Guo2020pra} if it satisfies both the unification condition above and~Eq.~\eqref{c1} holds for all $\rho\in\mS^{A_1A_2\cdots A_n}$ whenever $X_1|X_2|\cdots|X_{k}\succ^b Y_1|Y_2|\cdots |Y_{l}$ additionally.
For instance, $E_f^{(n)}$ is a complete multipartite entanglement measure
and there do exist unified multipartite entanglement measure but not complete~\cite{Guo2020pra}.
For the coarser relation of type (c), it is automatically true for any entanglement measure, i.e., Eq.~\eqref{c1} holds
for all states obey the coarser relation $\succ^c$, since the partial trace
is a specific LOCC (local operation and classical communication) and entanglement is non-increasing under LOCC.

With this scenario in mind, we now begin to investigate the completeness of the MQMI $I$ and $I'$.
It is clear that 
\beax
I(A_1:A_2:\cdots:A_n)=I(A_{\pi(1)}:A_{\pi(2)}:\cdots:A_{\pi(n)})
\eeax
and
\beax
I'(A_1:A_2:\cdots:A_n)=I'(A_{\pi(1)}:A_{\pi(2)}:\cdots:A_{\pi(n)})
\eeax
for any $\rho^{A_1A_2\cdots A_n}\in\mS^{A_1A_2\cdots A_n}$ and any permutation $\pi$.
Analogous to Eq.~\eqref{additivity},
we can prove that
\bea\label{additivity2}
I(A_1:A_2:\cdots:A_n)\quad\quad\quad\quad\quad\quad\quad\quad\quad\quad\quad\quad\quad\quad\nonumber\\
=I(A_1:\cdots:A_k)+I(A_{k+1}:\cdots:A_n)\quad \quad
\eea
whenever $I(A_1A_2\cdots A_k:A_{k+1}\cdots A_n)=0$, i.e., $\rho^{A_1A_2\cdots A_n}=\rho^{A_1A_2\cdots A_k}\otimes\rho^{A_{k+1}\cdots A_n}$ equivalently.
In fact
$I(A_1:A_2:\cdots :A_n)
=\sum\limits_{i=1}^nS_{A_i}-S_{A_1A_2\cdots A_n}=
\sum\limits_{i=1}^nS_{A_i}-S_{A_1A_2\cdots A_k}-S_{A_{k+1}:\cdots:A_n}=
I(A_1:\cdots:A_k)+I(A_{k+1}:\cdots:A_n)$,
which is straightforward, where $S_X:=S(X)$.
For $I'$, we take $n=4$ for example at first. If $I'(AB:CD)=0$ (i.e., $\rho^{ABCD}=\rho^{AB}\ot\rho^{CD}$), then
$I'(A:B:C:D)
=S_{ABC}+S_{BCD}+S_{ABD}+S_{ACD}-3S_{ABCD}=
[(S_{AB}+S_C)+(S_B+S_{CD})+(S_A+S_{CD})+(S_{AB}+S_D)]-3(S_{AB}+S_{CD})=
(S_A+S_B-S_{AB})+(S_C+S_D-S_{CD})=I'(A:B)+I'(C:D)
$. In general, we can get
\bea\label{additivity3}
I'(A_1:A_2:\cdots:A_n)\quad\quad\quad\quad\quad\quad\quad\quad\quad\quad\quad\quad\quad\quad\nonumber\\
=I'(A_1:\cdots:A_k)+I'(A_{k+1}:\cdots:A_n)\quad \quad
\eea
for any state with $I'(A_1A_2\cdots A_k:A_{k+1}\cdots A_n)=0$.
As that of Eq.~\eqref{additivity},
Eq.~\eqref{additivity2} and Eq.~\eqref{additivity3}
refer to that, if there is no mutual information between subsystem $A_1A_2\cdots A_k$
and subsystem $A_{k+1}\cdots A_n$, then the global mutual information contained only in
the system $A_1A_2\cdots A_k$ and the system $A_{k+1}\cdots A_n$ independently.
Henceforward, we call the measure is \textit{additive} if it satisfies the relation such as Eq.~\eqref{additivity},
Eq.~\eqref{additivity2} and Eq.~\eqref{additivity3}.
Namely, mutual information and complete multipartite entanglement measure are additive.

We now discuss whether $I$ and $I'$ are decreasing under coarsening of the system. Namely, whether the counterpart of Eq.~\eqref{c1} for $I$ and $I'$ are valid under the coarser relation $\succ^a$, $\succ^b$, and $\succ^c$. 

\begin{pro}\label{pro1} 	
	Let $X_1|X_2| \cdots |X_{k}$ and $Y_1|Y_2| \cdots |Y_{l}$ be two partitions of $A_1A_2\cdots A_n$ or subsystem of $A_1A_2\cdots A_n$. If $X_1|X_2| \cdots |X_{k} \succ Y_1|Y_2| \cdots |Y_{l}$, then 
\bea\label{c1'} 
I(X_1:X_2:\cdots:X_{k})\geqslant I(Y_1:Y_2:\cdots:Y_{l})
\eea
and
\bea\label{c2'} 
I'(X_1:X_2:\cdots:X_{k})\geqslant I'(Y_1:Y_2:\cdots:Y_{l})
\eea
hold for any
$\rho^{A_1A_2\cdots A_n}\in\mS^{A_1A_2\cdots A_n}$.
\end{pro}

\begin{proof}
	It is clear that
	\beax 
	&&I(A:B:C)-I(A:B)\\
	&=&(S_A+S_B+S_C-S_{ABC})-(S_A+S_B-S_{AB})\\
	&=&S_{AB}+S_C-S_{ABC}\geqslant 0
	\eeax
	since $S$ is subadditive, and 
		\beax 
	&&I'(A:B:C)-I'(A:B)\\
	&=&(S_{AB}+S_{BC}+S_{AC}-2S_{ABC})-(S_A+S_B-S_{AB})\\
	&=&(S_{AB}+S_{BC}-S_{ABC}-S_B)\\
	&&+(S_{AB}+S_{AC}-S_{ABC}-S_A)\\
	&\geqslant& 0
	\eeax
	since $S$ is strong subadditive (i.e., $S_{AB}+S_{BC}\geqslant S_{ABC}+S_B$ for any state).
	In general,
	\beax 
	&&I(A_1:A_2:\cdots A_n)-I(A_1:A_2:\cdots A_{n-1})\\
	&=&S_{A_1A_2\cdots A_{n-1}}+S_{A_n}-S_{A_1A_2\cdots A_n}\geqslant 0,
	\eeax
	and
	\beax 
	&&I'(A_1:A_2:\cdots A_n)-I'(A_1:A_2:\cdots A_{n-1})\\
	&=&\sum\limits_{i=1}^{n-1}\left( S_{\overline{A_i}}+S_{\overline{A_n}}-S_{A_1A_2\cdots A_n}-S_{\overline{A_iA_n}}\right) \\
	&\geqslant &0.
	\eeax
	That is, both $I$ and $I'$ are nonincreasing under the coarsening relation of
	type (a).
	
	Similarly, in light of the subadditivity and the strong subadditivity of the von Neumann entropy, we can obtain Eq.~\eqref{c1'} and Eq.~\eqref{c2'} under the
	coarsening relation of
	types (b) and (c).  
	For example, we can get $I'(AB:CD:EF)\geqslant I'(AB:C:E)$
	due to the subadditivity and the strong subadditivity of the von Neumann entropy.
	In fact, $I'(AB:CD:EF)-I'(AB:C:E)=
	(S_{ABCD}+S_{ABEF}+S_{CDEF}-2S_{ABCDEF})-(S_{ABC}+S_{ABE}+S_{CE}-2S_{ABCE})=(S_{ABCD}+S_{ABEF}+S_{CDEF}+2S_{ABCE})-(S_{ABC}+S_{ABE}+S_{CE}+2S_{ABCDEF})=[(S_{ABCD}+S_{ABCE})+(S_{ABEF}+S_{ABCE})+S_{CDEF}]-(S_{ABC}+S_{ABE}+S_{CE}+2S_{ABCDEF})\geqslant
	[(S_{ABC}+S_{ABCDE})+(S_{ABE}+S_{ABCEF})+S_{CDEF}]-(S_{ABC}+S_{ABE}+S_{CE}+2S_{ABCDEF})=
	[S_{ABCDE}+(S_{ABCEF}+S_{CDEF})]-(S_{CE}+2S_{ABCDEF})\geqslant
	[S_{ABCDE}+(S_{CEF}+S_{ABCDEF})]-(S_{CE}+2S_{ABCDEF})=
	(S_{ABCDE}+S_{CEF})-(S_{CE}+S_{ABCDEF})\geqslant0$.	
\end{proof}

Proposition~\ref{pro1} indicates that $I$ and $I'$ are well-defined complete 
measures in the sense of Refs.~\cite{Guo2020pra,G2021qst,Guo2022entropy}.
Henceforward, we call such a measure complete measure.
Under this framework, the mutual information between different subsystems can be compared with each other
in a clear hierarchic structure sense, from which we can discuss the distribution of the corresponding quantity 
thoroughly and comprehensively.

\section{Complete monogamy of $I$}

Having discussed the underlying concept of the complete MQMI, we now restrict attention to present the definition of the complete monogamy for MQMI with the same essence as that of 
the complete monogamy of the multipartie entanglement~\cite{Guo2020pra,Guo2022entropy} and the complete monogamy of the multipartite quantum discord~\cite{G2021qst}.
Let $J=I$ or $J=I'$. With the notations aforementioned, 
(i) we call $J$ is monogamous if it 
satisfies the \textit{dis-correlated condition}, i.e., 
for any state $\rho\in\mS^{ABC}$ that satisfies 
\bea\label{dis-correlated}
J(A:BC)=J(A:B)
\eea
we have that
\bea
J(A:C)=0.
\eea
(ii) $J$ is said to be completely monogamous if it 
	satisfies the \textit{complete dis-correlated condition}, i.e., for any state $\rho\in\mS^{A_1A_2\cdots A_n}$ that satisfies 
	\bea\label{dis-correlated2}
	J({X_1|X_2| \cdots|X_{k}})=J({Y_1|Y_2|\cdots|Y_{l}})
	\eea
	we have that
	\bea
	J({\Gamma})=0
	\eea
	holds for all $\Gamma\in \Xi(X_1|X_2| \cdots| X_{k}- Y_1|Y_2| \cdots |Y_{l})$, where $X_1|X_2| \cdots |X_{k}$ and $Y_1|Y_2| \cdots |Y_{l}$ are arbitrarily given partitions of $A_1A_2\cdots A_m$ or subsystem of $A_1A_2\cdots A_m$, and where $X_1|X_2| \cdots| X_{k}\succ^a Y_1|Y_2| \cdots |Y_{l}$.
(iii) $J$ is said to be tightly complete monogamous if we replace $\succ^a$ by $\succ^b$ in the above item (ii),
and the counterpart of Eq.~\eqref{dis-correlated2} is called \textit{tightly complete dis-correlated condition} instead.

In such a sense, according to the proof of Theorem 1 in Ref.~\cite{GG}, (i)
if $J$ is monogamous, then there exists $\alpha>0$ such that
\beax
J^\alpha(A:BC)\geqslant J^\alpha(A:B)+J^\alpha(A:C)
\eeax
holds for any state in $\mS^{ABC}$, where $\alpha$ is related to the dimension of $\mH^{ABC}$.
We observe here that $I$ and $I'$ are continuous fuctions since the von Neumann entropy is continuous. 
(ii) If $J$ is completely monogamous, then (we take $n=3$ for example)
\beax
J^\alpha(A:B:C)\geqslant J^\alpha(A:B)+J^\alpha(A:C)+J^\alpha(B:C)
\eeax
holds for any state in $\mS^{ABC}$ with $\alpha$ as above.
(iii) If $J$ is tightly complete monogamous, then 
\beax
J^\alpha(A:B:C)\geqslant J^\alpha(A:BC)+J^\alpha(B:C)
\eeax
holds for any state in $\mS^{ABC}$ for some $\alpha>0$ as above.
We are now ready for present the first main result of this article, which shows that
the MQMI $I$ is a nice measure of quantumness, but another alternative $I'$ is not since it is neither completely monogamous nor
tightly complete monogamous.

\begin{theorem}\label{monogamy-of-I}
	(i) $I$ is monogamous only on pure states.
	(ii) $I$ is not only completely monogamous but also
	tightly complete monogamous.
	(iii) $I'$ is neither completely monogamous nor
	tightly complete monogamous.
\end{theorem}

\begin{proof}
	
	(i) If $I(A:BC)=I(A:B)$, then $S_{AB}+S_{BC}-S_B-S_{ABC}=0$.
	According to Theorem 6 in Ref.~~\cite{HaydenJozaPetsWinter}, such a state $\rho^{ABC}$ is precisely the state that saturates the strong subadditivity of the von-Neumann entropy (i.e., the Markov state). For such states, the state space of system $B$, $\mH^B$, must have a decomposition into a direct sum of tensor products
	$
	\mH^B=\bigoplus_{j}\mH^{B_{j}^{L}}\otimes\mH^{B_{j}^{R}}
	$,
	such that $\rho^{ABC}$ admits the form
	\be\label{Markov}
	\rho^{ABC}=\bigoplus_{j}q_j\rho^{AB_{j}^{L}}\otimes\rho^{B_{j}^{R}C}\;,
	\ee
	where $q_j$ is a probability distribution~\cite{HaydenJozaPetsWinter}. 
	However, $\rho^{ABC}\neq\rho^{AB}\ot\rho^C$ whenever $\rho^{AB_{j}^{L}}\neq\rho^A\ot\rho^{B_{j}^{L}}$
	or $\rho^{B_{j}^{R}C}\neq \rho^{B_{j}^{R}}\ot\rho^{C}$.
	That is, $I(A:BC)=I(A:B)$ can not guarantee $I(A:C)=0$.

	(ii) If $I(A:B:C)=I(A:B)$, then $S_{AB}+S_C=S_{ABC}$,
	which implies that $\rho^{ABC}=\rho^{AB}\ot\rho^C$.
	Hence, $I(A:C)=I(B:C)=0$.
	If $I(A:B:C:D)=I(A:B:C)$, then $S_{ABC}+S_D=S_{ABCD}$, which leads to
	$\rho^{ABCD}=\rho^{ABC}\ot\rho^D$. Thus $I(A:D)=I(B:D)=I(C:D)=0$.
	Similarly,  $I(A:B:C:D)=I(A:B)$ implies $\rho^{ABCD}=\rho^{AB}\ot\rho^C\ot\rho^D$.
	In general, $I(A_1:A_2:\cdots:A_n)=I(A_{k_1}:A_{k_2}:\cdots A_{k_m})$ ($m<n$, $k_i\neq k_j$ whenever $i\neq j$, $1\leqslant k_i\leqslant n$) implies $\rho^{A_1A_2\cdots A_n}=\rho^{A_{k_1}A_{k_2}\cdots A_{k_m}}\ot\rho^{A_{k_{m+1}}}\ot\cdots\ot\rho^{A_{k_{n}}}$, and therefore
	$I(A_{k_1}A_{k_2}\cdots A_{k_m}:A_{k_{m+1}}:\cdots:A_{k_{n}})=0$.
	This yields $I(\Gamma)=0$ for any $\Gamma\in\Xi(A_1|A_2|\cdots|A_n-A_{k_1}|A_{k_2}|\cdots|A_{k_m})$.
	Namely, $I$ is completely monogamous.

	One can readily check that  
	$I(A:B:C:D)=I(A:BCD)$ implies $\rho^{BCD}$ is a product state (i.e., $I(B:C:D)=0$) and
	$I(A:B:C:D)=I(AB:CD)$ implies $\rho^{AB}$ and $\rho^{CD}$ are product states (i.e., $I(A:B)=I(C:D)=0$).
	In general, 
	\beax 
	&&I(A_1:A_2:\cdots:A_n)\\
	&=&I\left( A_{k^{(1)}_1}A_{k^{(1)}_2}\cdots A_{k^{(1)}_s}:A_{k^{(2)}_1}A_{k^{(2)}_2}\cdots A_{k^{(2)}_t}\right. \\
	&&~~\left. :\cdots:A_{k^{(l)}_1}A_{k^{(l)}_2}\cdots A_{k^{(l)}_r}\right) 
	\eeax
	 implies 
	 $I(A_{k^{(p)}_1}:A_{k^{(p)}_2}:\cdots:A_{k^{(p)}_q})=0$ for any $1\leqslant p\leqslant l$ and $q\in\{s, t, \dots, r\}$,
where $A_{k^{(1)}_1}\cdots A_{k^{(1)}_s}|A_{k^{(2)}_1}\cdots A_{k^{(2)}_t}|\cdots |A_{k^{(l)}_1}\cdots A_{k^{(l)}_r}$ is a $l$-partition of $A_1A_2\cdots A_n$ up to some permutation of the subsystems.
That is, $I$ is tightly complete monogamous.

(iii) We assume that $I'(A:B:C)=I(A:B)$,
	i.e.,
	$S(AB)+S(AC)+S(BC)-2S(ABC)=S(A)+S(B)-S(AB)$.
	Since
	$S(AB)+S(AC)\geqslant S(ABC)+S(A)$ and $S(AB)+S(BC)\geqslant S(ABC)+S(B)$
	we get
	$S(AB)+S(AC)= S(ABC)+S(A)$ and $S(AB)+S(BC)= S(ABC)+S(B)$.
	If
	the state Hilbert spaces $\mH^A$, $\mH^B$ have decompositions into a direct sum of tensor products as
	\[
	\mH^A=\bigoplus_{j}\mH^{A_{j}^{L}}\otimes\mH^{A_{j}^{R}},\; \mH^B=\bigoplus_{j}\mH^{B_{j}^{L}}\otimes\mH^{B_{j}^{R}},
	\] 
	such that 
	\bea
	\rho^{ABC}=\bigoplus_jq_j\rho_{a_j^L}\otimes\rho_{a_j^Rb_j^L}\otimes\rho_{b_j^R}\otimes\rho_{c_j},
	\eea
	it is easily checked that $S(AB)+S(AC)= S(ABC)+S(A)$ and $S(AB)+S(BC)= S(ABC)+S(B)$.
	But $I(B:C)>0$ and $I(A:C)>0$ provided that
	$\rho_{c_i}\neq\rho_{c_j}$ whenever $i\neq j$.
	Thus $I'$ is not completely monogamous.
	
	If $I'(A:B:C)=I(A:BC)$, we get
	$S(AB)+S(AC)=S(ABC)+S(A)$. That is, $\rho^{BAC}$ admits the form as Eq.~\eqref{Markov},
	which reveals that $\rho^{BC}$ is not necessarily a product state. Therefore
	$I'$ is not tightly complete monogamous.
	Together with Proposition~\ref{pro1}, the proof is completed. 	
\end{proof}

For the higher dimensional case, $I~(=I')$ is only monogamous on pure states since 
it is reduced to $2E_f$ for pure states and $E_f$ is monogamous~\cite{GG2019}.
Theorem~\ref{monogamy-of-I} indicates that complete monogamy does not imply monogamy in general although the measure
is a complete one.

In Ref.~\cite{GG}, we showed that the Markov quantum state satisfies the disentangling condition
\bea
E(A|BC)=E(AB)
\eea
for any bipartite entanglement monotone $E$. Hereafter, we always assume that $E$ is a bipartite entanglement monotone.
Thus, from the proof of Theorem~\ref{monogamy-of-I}, we have
\begin{itemize}
	\item[(i)] $I(A:BC)=I(A:B)$ implies $E(A|BC)=E(AB)$ and $E(C|AB)=E(CA)$;
	\item[(ii)] $I'(A:B:C)=I(A:B)$ implies $E(A|BC)=E(B|AC)=E(AB)$ and $E(C|AB)=E(CA)=E(CB)$.
\end{itemize}
In general, we can prove that
$$I'(A_1:A_2:\cdots:A_n)=I'(A_1:A_2:\cdots:A_k)$$ implies $$E(A_i|\overline{A_i})=E(A_i|\overline{A_iA_n})=E(A_i|\overline{A_iA_{i_1}A_{i_2}\cdots A_{i_l}A_n})$$ and $$E(A_n|\overline{A_n})=E(A_n|\overline{A_iA_n})=E(A_n|\overline{A_iA_{i_1}A_{i_2}\cdots A_{i_l}A_n})$$
for any $i_s\neq i$, $1\leqslant i\leqslant k$, $l<n-2$.
That is, the dis-correlated condition and the complete dis-correlated condition of MQMI
are closely related to the disentangling condition of entanglement.

\section{Complete monogamy of the MQMI via the Tsallis entopy}

In this section, we explore the mutual information deduced by the Tsallis entopy. 
The Tsallis $q$-entropy $S_q$ is defined by~\cite{Raggio}
\[S_q(\rho)=\frac{1-\tr\rho^q}{q-1},\quad q>0, ~q\neq1.\]
$S_q$ is subadditive when $q>1$~\cite{Audenaerta2007jmp},
i.e.,
\bea\label{sub-additive2}
S_q({AB})\leqslant S_q(A)+S_q(B),\quad q>1
\eea
for any $\rho^{AB}\in\mS^{AB}$,
where $\rho^{A,B}=\tr_{B,A}\rho^{AB}$.
Replace the von Neumann entropy with $S_q$ in Eqs.~\eqref{def-of-mui} and~\eqref{def-of-mui'}, we get
\bea\label{def-of-mui2}
I_q(A_1:A_2:\cdots :A_n)\quad\quad\quad\quad\quad\quad\quad\nonumber\\
:=\sum_{k=1}^nS_q({A_k})-S_q(A_1A_2\cdots A_n)
\eea
and
\bea\label{def-of-mui2'}
I'_q(A_1:A_2:\cdots :A_n)\quad\quad\quad\quad\quad\quad\quad\quad\quad\quad\quad\quad\nonumber\\
:=\sum_{k=1}^nS_q(\overline{A_k})
-(n-1)S_q(A_1A_2\cdots A_n),~~
\eea
respectively.
It is straightforward that
\bea\label{nonnegtive}
I_q(A_1:A_2:\cdots :A_n)\geqslant 0.
\eea
In Ref.~\cite{Guo2020pra}, we proved that, for any bipartite state $\rho_{AB}\in\mS^{AB}$,  $1+\tr\left(\rho^2_{AB}\right)=
\tr\left(\rho^2_A\right)+\tr\left(\rho^2_B\right)$ if and only if
$\rho_{AB}=\rho_A\ot\rho_B$ with
$\min\left\lbrace \rank\left( \rho_A\right) ,\rank\left( \rho_B\right) \right\rbrace =1$.
Thus, the equality for $q=2$ in Eq.~\eqref{nonnegtive} holds if and only if
$\rho^{A_1A_2\cdots A_n}$ is a product state with at most one of the reduced states $\rho^{A_i}$
has rank greater than 1.
$I_q'(A_1:A_2)=I_q(A_1:A_2)\geqslant 0$ for any bipartite state. However, 
\bea
I_q'(A_1:A_2:\cdots :A_n)\ngeqslant 0
\eea 
in general when $n>2$
due to the fact that $S_q$ is not strong subadditive~\cite{Petza2015},
i.e.,
\beax
S_q(AB)+S_q(BC)\ngeqslant S_q(ABC)+S_q(B),~~ q>0, ~q\neq1,
\eeax
in general.

Notice that there is another approach of MQMI, which is defind by (we take $n=3$ for example)~\cite{Chakrabarty2011,Horodecki2005,Kumar2017pra}
\bea\label{mui3'}
I''(A:B:C)\quad\quad\quad\quad\quad\quad\quad\quad\quad\quad\quad\quad\quad\quad\quad\quad\quad\quad\nonumber\\
=S_A+S_B+S_C-(S_{AB}+S_{AC}+S_{BC})+S_{ABC}.~~\quad
\eea
$I''$ can be negative~\cite{Chakrabarty2011,Horodecki2005}
and thus it is not a good alternative of MQMI.
We now consider this quantity by replacing $S$ with $S_q$, i.e.,
\bea\label{def-of-mui3'}
I''_q(A:B:C)
=S_q(A)+S_q(B)+S_q(C)\quad\quad\quad\quad\quad\quad\nonumber\\
-[S_q({AB})+S_q({AC})+S_q({BC})]+S_q({ABC}).~~\quad
\eea
Take the three qubit state $\rho=\frac12|{\rm GHZ}\ra\la{\rm GHZ}|+\frac{1}{16}I$,
one can easily get $I''_q(A:B:C)<0$.
Namely, this approach is not valid for the Tsallis $q$-entropy MQMI, either.

By definition, $I_q$ and $I_q'$ are symmetric under permutation of the subsystems.
We next show that $I_q$ is additive while $I_q'$ is not.
If $I_q(A_1A_2\cdots A_k:A_{k+1}\cdots A_n)=0$,
then
$S_q(A_1A_2\cdots A_k)+S_q(A_{k+1}\cdots A_n)=S_q(A_1A_2\cdots A_n)$,
which yields
\bea
I_q(A_1:A_2:\cdots:A_n)\quad\quad\quad\quad\quad\quad\quad\quad\quad\quad\quad\quad\quad\quad\nonumber\\
=I_q(A_1:A_2:\cdots:A_k)+I_q(A_{k+1}:\cdots:A_n).\quad\quad
\eea
But $I_q'$ does not obey such a equality.
In order to see this,
we take $\rho^{ABCD}=\rho^{ABC}\ot\rho^D$ with
\beax 
\rho^{ABC}=
\left( \begin{array}{cccccccc}
0&0&0&0&0&0&0&0\\
0&0&0&0&0&0&0&0\\
0&0&\frac14&0&\frac14&0&0&0\\
0&0&0&\frac14&0&\frac14&0&0\\
0&0&\frac14&0&\frac14&0&0&0\\
0&0&0&\frac14&0&\frac14&0&0\\
0&0&0&0&0&0&0&0\\
0&0&0&0&0&0&0&0\\	
\end{array}\right),~~\rho^{D}= \left( \begin{array}{cc}
\frac12&0\\
0&\frac12
\end{array}\right).
\eeax
It follows that 
\beax 
\rho^{AB}&=&
\left( \begin{array}{cccc}
	0&0&0&0\\
	0&\frac12&\frac12&0\\
	0&\frac12&\frac12&0\\
	0&0&0&0
\end{array}\right),\\
\rho^{AC}&=&\rho^{BC}=\rho^{CD}=
\left( \begin{array}{cccc}
\frac14&0&0&0\\
0&\frac14&0&0\\
0&0&\frac14&0\\
0&0&0&\frac14
\end{array}\right).
\eeax
Short computation 
gives
$I_q'(AB:CD)=S_q(AB)+S_q(CD)-S_q(ABCD)=S_q(CD)-S_q(ABCD)=\frac{1}{q-1}(1-4^{1-q})-\frac{1}{q-1}(1-4^{1-q})=0$,
but
$\frac{1}{q-1}\{I_q'(A:B:C:D)-[I_q(A:B)+I_q(C:D)]\}
=\frac{1}{q-1}\{[(1-2^{1-q})+(1-8^{1-q})+(1-2^{1-q})+(1-8^{1-q})-3(1-4^{1-q})]
-[4(1-2^{1-q})-(1-4^{1-q})]\}
=\frac{1}{q-1}(4^{1-q}+2^{1-q}-1-8^{1-q})=\frac{1}{q-1}(2^{1-q}-1)(1-4^{1-q})>0$ for any $q>1$.
From the subadditivity of $S_q$, the following is straightforward.

\begin{pro}\label{pro2} 	
	Let $X_1|X_2| \cdots |X_{k}$ and $Y_1|Y_2| \cdots |Y_{l}$ be two partitions of $A_1A_2\cdots A_n$ or subsystem of $A_1A_2\cdots A_n$. If $X_1|X_2| \cdots |X_{k} \succ^{a,b} Y_1|Y_2| \cdots |Y_{l}$, then 
	\bea\label{c1''} 
	I_q(X_1:X_2:\cdots:X_{k})\geqslant I_q(Y_1:Y_2:\cdots:Y_{l})
	\eea
	hold for any
	$\rho^{A_1A_2\cdots A_n}\in\mS^{A_1A_2\cdots A_n}$.
\end{pro}

However, one can readily check that $I_q$ may increase under the coarsening relation of type (c).
That is, $I_q$ displays some degree of `completeness' as a multipartite measure 
but $I'_q$ fails.

We next explore the complete monogamy and monogamy of $I_q$.
Let $X_1|X_2| \cdots |X_{k}$ and $Y_1|Y_2| \cdots |Y_{l}$ be two partitions of $A_1A_2\cdots A_n$ or subsystem of $A_1A_2\cdots A_n$. If $X_1|X_2| \cdots |X_{k} \succ^{a,b} Y_1|Y_2| \cdots |Y_{l}$, and
\beax
I_q(X_1:X_2:\cdots:X_{k})=I_q(Y_1:Y_2:\cdots:Y_{l}),
\eeax
we can easily get 
\bea
I_q({\Gamma})=0
\eea
holds for all $\Gamma\in \Xi(X_1|X_2| \cdots| X_{k}- Y_1|Y_2| \cdots |Y_{l})$.
In particular, for $q=2$, $\Gamma$ is a product state with at most one of the reduced states has rank greater than 1.
We thus get the following result.

\begin{theorem}\label{monogamy-of-I_q}
	(i) $I_q$ is monogamous on pure states.
	(ii) $I_q$ is not only completely monogamous but also
	tightly complete monogamous under the coarsening relation of types (a) and (b).
\end{theorem}

\begin{proof}
We only need to check item (i). For pure state, $I_q$ reduces to the Tsallis $q$-entropy of entanglement, where 
the Tsallis $q$-entropy of entanglement is defined by~\cite{Guo2020pra}
\beax 
E^{(n)}_q(|\psi\ra)=\frac12\left[S_q(A_1)+S_q(A_2)+\cdots+S_q(A_n)\right], ~q>1
\eeax
for pure state $|\psi\ra\in\mH^{ABC}$, 
and then define by the convex-roof extension for mixed states. Thus $I_q$ is monogamous on pure states since the Tsallis $q$-entropy of entanglement is monogamous~\cite{GG2019}.
\end{proof}

It is worth mentioning that $I_q$ is monogamous iff $S_q(AB)+S_q(BC)=S_q(ABC)+S_q(B)$
implies $S_q(AC)=S_q(A)+S_q(C)$. We remark here that this is not true. Taking~\cite{Chengshuming} 
\bea
\rho^{ABC}=p|000\ra\la000|+(1-p)|111\ra\la111|,
\eea 
we get $S_q(AB)+S_q(BC)=S_q(ABC)+S_q(B)$ but $S_q(AC)<S_q(A)+S_q(C)$.
Comparing with $I$ and $I'$, as a measure of mutual information, 
$I_q$ is nicer than $I'_q$ but worse than $I$. 
For $0<q<1$, $S_q$ is neither subadditive nor superadditive~\cite{Raggio}
(superadditive refers to $S_q(AB)\geqslant S_q(A)+S_q(B)$),
so we can not define the associated mutual information whenever $0<q<1$. The R\'{e}nyi $\alpha$-entropy (i.e., $R_{\alpha}(\rho):=(1-\alpha)^{-1}\ln(\tr\rho^\alpha)$, $0<\alpha< 1$)
is the same since it is not subadditive~\cite{Aczel}, either.
We call $I$ and $I_q$ the type-$1$ MQMI, $I'$ and $I_q'$ the type-$2$ MQMI, $I''$ and $I_q''$ the type-$3$ MQMI.
Together with Proposition~\ref{pro1} and Theorem~\ref{monogamy-of-I}, we find out that
the typ-1 is nicer than the typ-2 for characterizing the mutual information as a measure of multipartite, and the type-3 can not be an alternative indeed.

Let us further remark that, for $q=2$, $S_q$ is the linear entropy, which can be regarded as a measure of purity~\cite{Fano}. Hence, MQMI is indeed a measure of the multipartite ``mutual purity'' and they are the same in nature.

\begin{table*}
	\caption{\label{tab:table1} Comparing of  $I$, $I'$, $I''$, $I_q$, $I_q'$, and $I_q''$.
		M, CM, TCM, and TI signify the measure is monogamous, completely monogamous, tightly completel monogamous, and satisfies the triangle inequality, respectively. ``$\succ^{a,b,c}$'' denotes the MQMI is non-increasing under the coarsening relation ``$\succ^{a,b,c}$''. ``$--$'' means the item is senseless or unknown.}	
	\begin{ruledtabular}
		\begin{tabular}{cccccccccccc}
			MQMI    & entropy& non-negative            & symmetric   & additivity  & $\succ^a$ &$\succ^b$ &$\succ^c$ & M & CM & TCM  & TI  \\ \colrule
			$I$    &$S$ &$\checkmark$         &$\checkmark$ &$\checkmark$&$\checkmark$&$\checkmark$ &$\checkmark$&pure states&$\checkmark$& $\checkmark$&$\checkmark$\\
			$I'$ &$S$ &$\checkmark$   &$\checkmark$ &$\checkmark$&$\checkmark$&$\checkmark$ &$\checkmark$ & pure states&$\times$ & $\times$ &$\times$\\
			$I''$  &$S$&$\times$         &$\checkmark$& $--$           & $--$    & $--$   & $--$  & $--$   &$--$ &$--$&$--$\\
			$I_q$  &$S_q$, $q>1$&$\checkmark$     &$\checkmark$&$\checkmark$&$\checkmark$&$\checkmark$ &$\times$  &pure states&$\checkmark\footnotemark[1]$&$\checkmark\footnotemark[2]$&$\times$\\
			$I_q'$  &$S_q$, $q>1$&$\times$      &$\checkmark$&       $\times$     &  $\times$   &$\times$   & $\times$  &pure states   &$\times$ &$\times$&$\times$\\
			$I_q''$  &$S_q$, $q>1$&$\times$                                &$\checkmark$          &$--$&   $--$         &  $--$   &  $--$  &   $--$& $--$  &$--$ &$--$
		\end{tabular}
	\end{ruledtabular}
	\footnotetext[1]{It is completely monogamous under the coarser relation $\succ^{a,b}$.}
	\footnotetext[2]{It is tightly complete monogamous under the coarser relation $\succ^{a,b}$.}
\end{table*}

\section{Triangle relation of QMI}

The first triangle relation for entanglement is the concurrence triangle for three-qubit pure state~\cite{Qian2018pra,Zhu2015pra}:
\bea
C^2_{A|BC}\leqslant C^2_{AB|C}+C^2_{B|AC}.
\eea
Very recently, we show that such a triangle relation is generally true~\cite{Guo2022jpa}.
Let $E$ be a continuous bipartite entanglement measure.
Then there exists
$0<\alpha<\infty$ such that~\cite{Guo2022jpa}
\bea\label{th1-in-2022jpa}
E^{\alpha}(|{A|BC})\leqslant  E^{\alpha}({B|AC})
+ E^{\alpha}({AB|C})
\eea
for all pure states $|\psi\ra^{ABC}\in\mathcal{H}^{ABC}$ with fixed $\dim\mH^{ABC}=d<\infty$.
	Let $E^{(3)}$ be a continuous unified tripartite entanglement measure.
	Then~\cite{Guo2022jpa}	
	\be\label{tetra-condition-1}
	E^{\alpha}({A|B|CD})\leqslant  E^{\alpha}({A|BD|C})
	+ E^{\alpha}({AD|B|C})
	\ee
	for all $|\psi\ra^{ABCD}\in\mathcal{H}^{ABCD}$ with $\alpha$ as above.
	Here we omit the superscript $^{(3)}$ of $E^{(3)}$ for brevity.
	Let $E$ be a continuous bipartite entanglement measure that is determined by the eigenvalues of the reduced state.
Then~\cite{Guo2022jpa}	
\be\label{tetra-condition-2}
E^{\alpha}({AB|CD})\leqslant  E^{\alpha}({AC|BD})
+ E^{\alpha}({AD|BC})
\ee
for all $|\psi\ra^{ABCD}\in\mathcal{H}^{ABCD}$ with $\alpha$ as above.
For mutual information $I$ and $I'$, we have the 
triangle relation below analogously.

\begin{pro}\label{triangle-relation}
The MQMI $I$ and $I'$ admit the following triangle relations
\bea\label{triangle-relation1}
I(A:BC)\leqslant I(B:AC)+I(AB:C)
\eea
for any state in $\mS^{ABC}$, 
and 
\bea\label{triangle-relation2}
I(AB:CD)\leqslant I(AC:BD)+I(AD:BC),\quad\quad\quad\quad \\
I(A:B:CD)\leqslant I(A:BD:C)+I(AD:B:C),\quad\quad\\
I'(A:B:CD)\leqslant I'(A:BD:C)+I'(AD:B:C)\quad\quad
\eea
hold for any state in $\mS^{ABCD}$, 
but
$I_q$ fails.
\end{pro}

\begin{proof}
It is easy to derive from the subadditivity and the strong subadditivity of the von Neumann entropy that
$I(B:AC)+I(AB:C)-I(A:BC)
=(S_B+S_{AC}-S_{ABC})+(S_{AB}+S_{C}-S_{ABC})
-(S_A+S_{BC}-S_{ABC})
=(S_{AB}+S_{AC}-S_{ABC}-S_A)+(S_B+S_{C}-S_{BC})
\geqslant0$,
$I(AC:BD)+I(AD:BC)-I(AB:CD)
=(S_{AC}+S_{BC}+S_{AD}+S_{BD})
-(S_{AB}+S_{CD}+S_{ABCD})
\geqslant(S_{ABC}+S_{C}+S_{ABD}+S_{D})
-(S_{AB}+S_{CD}+S_{ABCD})
=(S_{ABC}+S_{ABD}-S_{ABCD}-S_{AB})
+(S_{C}+S_{D}-S_{CD})
\geqslant0$,
$I(A:BD:C)+I(AD:B:C)-I(A:B:CD)
=(S_{AD}+S_{BD}+2S_{C})
-(S_{ABCD}+S_{CD})
\geqslant(S_{ABD}+S_{D}+2S_{C})
-(S_{ABCD}+S_{CD})
\geqslant(S_{ABD}+S_{CD}+S_{C})
-(S_{ABCD}+S_{CD})
=S_{ABD}+S_{C}
-S_{ABCD}
\geqslant0$,
and
$I'(A:BD:C)+I'(AD:B:C)-I'(A:B:CD)
=(2S_{ABD}+S_{AC}+S_{BC})
-(2S_{ABCD}+S_{AB})
\geqslant(S_{ABCD}+S_{A}+S_{ABD}+S_{BC})
-(2S_{ABCD}+S_{AB})
\geqslant S_{ABC}+S_{ABD}
-S_{ABCD}-S_{AB}
\geqslant0$.
For $I_q$, by the invalidation of the strong subadditivity of the Tsallis $q$-entropy, 
the proof is completed.
\end{proof}

That is, the von Neumann entropy MQMI reflects the same triangle relation as that of entanglement. We now also conclude that the von Neumann entropy MQMI sounds nicer than that of Tsallis entropy.
For more clarity, we list all the properties of these measures so far in Table~\ref{tab:table1}.
We close this section with
the following inequalities which reveal the relation between entanglement and 
the mutual information.

\begin{pro} Let $\rho$ be any state in $\mS^{A_1A_2\cdots A_n}$. Then
\bea
I(\rho)+S(\rho)\geqslant2E_f^{(n)}(\rho)
\eea
and 
\bea
I_q(\rho)+S_q(\rho)\geqslant2E_q^{(n)}(\rho),
\eea
and the equality holds iff $\rho$ is a pure state.
\end{pro}

\begin{proof}
We assume with no loss of generality that $n=3$.
For any given $\rho\in\mS^{ABC}$,
let 
$E_f^{(3)}(\rho)=\sum\limits_{i}p_i[E_f^{(3)}(|\psi_i\ra\la\psi_i|)]
=\frac12\sum\limits_{i}p_i[S(\rho_i^{A})+S(\rho_i^{B})+S(\rho_i^{C})]$,
where $\rho_i^X=\tr_{\bar{X}}|\psi_i\ra\la\psi_i|$.
It follows that
$I(\rho)+S(\rho)-2E_f^{(n)}(\rho)
=S(\rho^A)+S(\rho^B)+S(\rho^C)
-\sum\limits_{i}p_i[S(\rho_i^{A})+S(\rho_i^{B})+S(\rho_i^{C})]
=S(\rho^A)-\sum\limits_{i}p_iS(\rho_i^{A})+S(\rho^B)-\sum\limits_{i}p_iS(\rho_i^{B})
+S(\rho^C)-\sum\limits_{i}p_iS(\rho_i^{C})
\geqslant0$
since $S$ is concave.
Notice additionally that $S$ is strictly concave~\cite{Wehrl1978},
the equality is immediate.
Applying the same strategy for the Tsallis $q$-entropy, we get the second
inequality and the equality makes sense for pure states in the light of the strict concavity of the Tsallis $q$-entropy. 
\end{proof}

That is, the sum of mutual information and the total entropy acts as a upper bound of entanglement.
It can be interpreted as the QMI referring more quantum correlation than entanglement.
We also need to note is that $E_f^{(n)}(\rho)\geq S(\rho)$ and $E_q^{(n)}(\rho)\geq S(\rho)$ for pure states
but $E_f^{(n)}(\rho)<S(\rho)$ and $E_q^{(n)}(\rho)<S(\rho)$ for any separable mixed state, namely, entanglement and the global entropy are incomparable.

\section{Conclusions and Discussions}

The completeness is a basic requirement for a multipartite measure. 
We have shown that the two types of MQMI via the standard von Neumann entropy
are complete measures, and the type-1 MQMI via the Tsallis $q$-entropy demonstrates some weak completeness while
the type-2 MQMI via the Tsallis $q$-entropy is not complete any more.
Moreover, we proved that the type-1 MQMI is not only completely monogamous but also tightly complete monogamous, but the type-2 MQMI fails.
We also found that the von Neumann entropy MQMI obeys the triangle relation which is 
the same as that of entanglement measure.

We thus conclude that
the von Neumann entropy is better than all the other versions of entropy from such a point of view, and
the type-1 von Neumann entropy MQMI represents the same quality as that of the complete measure of multipartite entanglement since both of them are complete and completely monogamous.
Together with multipartite entanglement measure and other measure of multipartite quantum correlations, we can get a better understanding the structure of the multipartite quantum correlation.

From the argument in the previous sections,
one can conclude that any non-negative quantity defined on quantum state can induce a corresponding MQMI provided it is subadditive.
Moreover, if it is strong subadditive additionally, it can then define a complete MQMI and is also completely monogamous and tightly complete monogamous in general for the type-1. 
Despite the type-1 MQMI we proposed is completely monogamous and tightly complete monogamous,
it is monogamous only on pure states. This indicates that
monogamy and complete monogamy may be independent on each other.
But it remains a open problem whether there exists a 
MQMI which is not only completely monogamous and tightly complete monogamous but also monogamous.

\begin{acknowledgements}
We are grateful to S.-M. Cheng for helpful
discussions. This work is supported by the National Natural Science Foundation of
China under Grant No.~11971277, the Fund Program for the Scientific Activities of Selected
Returned Overseas Professionals in Shanxi Province under Grant No.~20220031, and the Scientific Innovation Foundation of the Higher
Education Institutions of Shanxi Province under Grant No.~2019KJ034.
\end{acknowledgements}



\begin{thebibliography} {99}

\bibitem{Nielsen} M.~A.~Nielsen, I.~L.~Chuang, \textit{Quantum Computatation and
Quantum Information}, (Cambridge University Press, Cambridge, 2000).



\bibitem{Wiseman2007prl} H. M. Wiseman, S. J. Jones, and A. C. Doherty, 
Steering, entanglement, nonlocality, and the Einstein-Podolsky-Rosen paradox,
Phys. Rev. Lett. \textbf{98}, 140402 (2007). 

\bibitem{Wiseman2007pra} S. J. Jones, H. M. Wiseman, and A. C. Doherty, 
Entanglement, Einstein-Podolsky-Rosen correlations, Bell nonlocality, and steering,
Phys. Rev. A \textbf{76}, 052116 (2007).


\bibitem{Ollivier2001quantum} H. Ollivier, and W. H. Zurek,  
Quantum discord: a measure of the quantumness of correlations,
{Phys. Rev. Lett.} \textbf{88}, 017901 (2001).

\bibitem{Henderson2001classical} L. Henderson, and V. Vedral,
Classical, quantum and total correlations,
{J. Phys. A: Math. Theor.} \textbf{34}, 6899 (2001).

\bibitem{Bennett1993teleporting} C. H. Bennett, G. Brassard, C. Cr\'{e}peau, R. Jozsa,
A. Peres, and W. K. Wootters, 
Teleporting an unknown quantum state via dual classical and Einstein-Podolsky-Rosen channels,
Phys. Rev. Lett. \textbf{70}, 1895 (1993).

\bibitem{Horodecki2009} R. Horodecki, P. Horodecki, M. Horodecki, and K. Horodecki,
Quantum entanglement,
Rev. Mod. Phys. \textbf{81}, 865 (2009).

\bibitem{Guhne2009} O. G\"{u}hne and G. T\'{o}th, 
Entanglement detection, 
Phys. Rep. \textbf{474},1 (2009).


\bibitem{Zhang2006experimental} Q. Zhang, A. Goebel, C. Wagenknecht, Y.-A. Chen,
B. Zhao, T. Yang, A. Mair, J. Schmiedmayer, and J.-W. Pan, 
Experimental quantum teleportation of a two-qubit composite system,
Nat. Phys. \textbf{2}, 678 (2006).



\bibitem{Bennett1992communication} C. H. Bennett and S. J. Wiesner, 
Communication via one-and two-particle operators on Einstein-Podolsky-Rosen states,
Phys. Rev. Lett. \textbf{69}, 2881 (1992).





\bibitem{Bennett1996prl} C.~H. Bennett, D.~P.~DiVincenzo, J.~A.~Smolin, and W.~K.~Wootters,
Mixed-state entanglement and quantum error correction,
Phys. Rev. A \textbf{54}, 3824~(1996).



\bibitem{Datta2008prl} A. Datta, A. Shaji, and C. M. Caves, 
Quantum discord and the power of one qubit,
Phys. Rev. Lett. \textbf{100}, 050502
(2008).


\bibitem{Verstraete2003pra} F. Verstraete, J. Dehaene, and B. D. Moor,
Normal forms and entanglement measures for multipartite quantum states,
Phys. Rev. A \textbf{68}, 012103 (2003).

\bibitem{Luque2003pra} J.-G. Luque and J.-Y. Thibon,
Polynomial invariants of four qubits,
Phys. Rev. A \textbf{67}, 042303 (2003).


\bibitem{Gour2010prl} G. Gour,
Evolution and symmetry of multipartite entanglement,
Phys. Rev. Lett. \textbf{105}, 190504 (2010).

\bibitem{Viehmann2011pra} O. Viehmann, C. Eltschka, and J. Siewert,
Polynomial invariants for discrimination and classification of four-qubit entanglement,
Phys. Rev. A \textbf{83}, 052330 (2011).

\bibitem{Osterloh2009jmp} A. Osterloh,  
On polynomial invariants of several qubits,
J. Math. Phys. \textbf{50}(3), 033509-033509 (2009).

\bibitem{Szalay} S. Szalay, Multipartite entanglement measures,
Phys. Rev. A  \textbf{92}, 042329 (2015).

\bibitem{Rulli2011pra} C. C. Rulli and M. S. Sarandy, 
Global quantum discord in multipartite systems
{Phys. Rev. A} \textbf{84}, 042109 (2011).		

\bibitem{Giorgi2011prl} G. L. Giorgi, B. Bellomo, F. Galve, \textit{et al.}, 
Genuine quantum and classical correlations in multipartite Systems,
{Phys. Rev. Lett.} \textbf{107}, 190501 (2011).


\bibitem{Radhakrishman2020prl} C. Radhakrishnan, M. Lauri{\`e}re, and T. Byrnes,
Multipartite generalization of quantum discord,
{Phys. Rev. Lett.} \textbf{124}, 110401 (2020).

\bibitem{Watanabe} S. Watanabe, 
Information theoretical analysis of multivariate correlation,
IBM J. Res. Dev. \textbf{4}, 66 (1960).

\bibitem{Kumar2017pra} A. Kumar, 
Multiparty quantum mutual information: An alternative definition,
Phys. Rev. A \textbf{96}, 012332 (2017).







\bibitem{Terhal2004} B. Terhal,
Is entanglement monogamous?
IBM J. Res. Dev. \textbf{48}, 71 (2004).

\bibitem{Bennett2014} C. H. Bennett, 
in \textit{Proceedings of the FQXi 4th International Conference, Vieques Island, Puerto Rico, 2014}, 
http://fqxi.org/conference/talks/2014.

\bibitem{Toner} B. Toner, 
Monogamy of non-local quantum correlations,
Proc. R. Soc. A \textbf{465}, 59 (2009).

\bibitem{Seevinck} M. P. Seevinck, 
Monogamy of correlations versus monogamy of entanglement,
Quantum Inf. Process. \textbf{9}, 273 (2010).

\bibitem{streltsov2012are} A. Streltsov, G. Adesso, M. Piani, D. Bru\ss, 
Are general quantum correlations monogamous? 
Phys. Rev. Lett. \textbf{109}, 050503 (2012).

\bibitem{Augusiak2014pra} R. Augusiak, M. Demianowicz, M. Paw\l{}owski, J. Tura, and A. Ac\'{\i}n,
Elemental and tight monogamy relations in nonsignaling theories,
Phys. Rev. A \textbf{90}, 052323 (2014).

\bibitem{Ma2011} X.-s. Ma, B. Dakic, W. Naylor, A. Zeilinger, and P.Walther,
Quantum simulation of the wavefunction to probe frustrated Heisenberg spin systems,
Nat. Phys. \textbf{7}, 399 (2011).

\bibitem{Brandao2013} F. G. S. L. Brandao and A.W. Harrow, in \textit{Proceedings
of the 45th Annual ACM Symposium on Theory of Computing,
2013}, http://dl.acm.org/citation.cfm?doid=2488608.2488718.

\bibitem{Garcia} A. Garc\'{\i}a-S\'{a}ez and J. I. Latorre,
Renormalization group contraction of tensor networks in three dimensions, 
Phys. Rev. B \textbf{87}, 085130 (2013).

\bibitem{Lloyd} S. Lloyd and J. Preskill, 
Unitarity of black hole evaporation in final-state projection models,
J. High Energy Phys. \textbf{08}, 126 (2014).

\bibitem{Pawlowski} M. Paw\l owski, 
Security proof for cryptographic protocols based only on the monogamy of Bell's inequality violations,
Phys. Rev. A \textbf{82}, 032313 (2010).

\bibitem{Coffman} V.~Coffman, J.~Kundu, and W.~K.~Wootters,
Distributed entanglement,
Phys. Rev. A \textbf{61}, 052306~(2000).

\bibitem{Koashi} M. Koashi and A. Winter,
Monogamy of quantum entanglement and other correlations,
Phys. Rev. A \textbf{69}, 022309 (2004).

\bibitem{Osborne} T. J. Osborne and F. Verstraete,
General monogamy inequality for bipartite qubit entanglement,
Phys. Rev. Lett. \textbf{96}, 220503 (2006).



\bibitem{Bai} Y.-K. Bai, Y.-F. Xu, and Z. D. Wang,
General monogamy relation for the entanglement of formation in multiqubit systems,
Phys. Rev. Lett. \textbf{113}, 100503 (2014).



\bibitem{Luo2016pra} Y. Luo, T. Tian, L.-H. Shao, and Y. Li,
General monogamy of Tsallis $q$-entropy entanglement in multiqubit systems,
Phys. Rev. A \textbf{93}, 062340 (2016).

\bibitem{Dhar} H. S. Dhar, A. K. Pal, D. Rakshit, A. S. De, and U Sen,
Monogamy of quantum correlations-a review,
In Lectures on General Quantum Correlations and their Applications, pp. 23-64. Springer, Cham, 2017.

\bibitem{Hehuan} H. He and G. Vidal,
Disentangling theorem and monogamy for entanglement negativity,
Phys. Rev. A \textbf{91}, 012339 (2015).

\bibitem{GG} G. Gour and Y. Guo, Monogamy of entanglement without inequalities,
Quantum \textbf{2}, 81 (2018).

\bibitem{GG2019} Y. Guo and G. Gour, Monogamy of the entanglement of formation,
Phys. Rev. A \textbf{99}, 042305 (2019).

\bibitem{Guo2020pra} Y. Guo and L. Zhang, Multipartite entanglement measure and complete monogamy relation,
Phys. Rev. A \textbf{101}, 032301 (2020).

\bibitem{G2021qst} Y. Guo, L. Huang, and Y. Zhang, 
Monogamy of quantum discord, 
Quant. Sci. Tech. \textbf{6}, 045028 (2021).

\bibitem{Guo2022entropy} Y. Guo, 
When is a genuine multipartite entanglement measure monogamous? 
Entropy \textbf{24}, 355 (2022).

\bibitem{Groisman2005pra} B. Groisman, S. Popescu, and A. Winter, 
Quantum, classical, and total amount of correlations in a quantum state,
Phys. Rev. A \textbf{72}, 032317 (2005).

\bibitem{Guo2014srep} Y. Guo and S. Wu, 
Quantum correlation exists in any non-product state, 
Sci. Rep. \textbf{4}, 7179 (2014).

\bibitem{Guo2015ijtp} Y. Guo, X. Li, B. Li, \textit{et al}., 
Quantum correlation induced by the average distance between the reduced states, 
Int. J. Theor. Phys. \textbf{54}, 2022-2030 (2015).


\bibitem{HaydenJozaPetsWinter} P. Hayden, R. Jozsa, D. Petz, and A. Winter,
Structure of states which satisfy strong subadditivity of quantum entropy with equality,
Commun. Math. Phys. \textbf{246}~(2), 359-374 (2004).


\bibitem{Raggio} G. A. Raggio, 
Properties of qentropies,
J. Math. Phys. \textbf{36}, 4785 (1995).


\bibitem{Audenaerta2007jmp} K. M. R. Audenaerta,
Sub additivity of $q$-entropies for $q>1$,
J. Math. Phys. \textbf{48}, 083507 (2007).


\bibitem{Petza2015} D. Petza and D. Virosztek, 
Some inequalities for quantum Tsallis entropy related to the strong subadditivity, 
arXiv: 1403.7062v3.


\bibitem{Chakrabarty2011} I. Chakrabarty, P. Agrawal, and A. K. Pati, 
Quantum dissension: Generalizing quantum discord for three-qubit states,
Eur. Phys. J. D \textbf{65},
605 (2011).


\bibitem{Horodecki2005} M. Horodecki, J. Oppenheim, and A. Winter, 
Partial quantum information,
Nature (London) \textbf{436}, 673 (2005).

\bibitem{Chengshuming} X. Ge and S.-M. Cheng, in preparation.

\bibitem{Aczel} J. Acz\'{e}l and Z. Dar\'{o}czy, On Measures of Information
and their Characterization, Academic Press, 1975.

\bibitem{Fano} U. Fano, 
Description of States in Quantum Mechanics by Density Matrix and Operator Techniques,
Rev. Mod. Phys. \textbf{29}, 74 (1957).

\bibitem{Qian2018pra} X. F. Qian, M. A. Alonso, and J. H. Eberly, 
Entanglement polygon inequality in qubit systems,
{New J. Phys.} \textbf{20}, 063012 (2018).

\bibitem{Zhu2015pra} X. N. Zhu and S. M. Fei, 
Generalized monogamy relations of concurrence for $N$-qubit systems,
{Phys. Rev. A} \textbf{92}, 062345 (2015).

\bibitem{Guo2022jpa} Y. Guo, Y. Jia, X. Li, and L. Huang, 
Genuine multipartite entanglement measure,
J. Phys. A: Math. Theor. \textbf{55}, 145303 (2022).

































\bibitem{Wehrl1978} A. Wehrl, 
General properties of entropy, 
Rev. Mod. Phys. \textbf{50},
221 (1978).







\end{thebibliography}


\end{document}